\documentclass[preprint,showpacs,preprintnumbers,amsmath,amssymb]{revtex4}
\usepackage{graphicx}
\usepackage{dcolumn}
\usepackage{bm}

\usepackage{color}
\definecolor{darkgreen}{rgb}{0,.5,0}
\usepackage[colorlinks,filecolor=blue,citecolor=red, urlcolor=darkgreen]{hyperref}

\begin{document}

\title{Unstable nuclei in dissociation of light stable and radioactive nuclei in nuclear track emulsion}

\author{D. A. Artemenkov$^1$, A. A. Zaitsev$^{1,2}$, P. I. Zarubin$^{1,2}$}
  \email{zarubin@lhe.jinr.ru}
\affiliation {${}^{1}$V.I. Veksler and A.M. Baldin Laborotory of High Energy Physics, \\Joint Institute for Nuclear Research, Russia\\
	${}^{2}$P. N. Lebedev Physical Institute of the Russian Academy of Sciences, Russia}

\begin{abstract}

A role of the unstable nuclei ${}^{6}$Be, ${}^{8}$Be and ${}^{9}$B in the dissociation of relativistic nuclei ${}^{7,9}$Be, ${}^{10}$B and ${}^{10,11}$C is under study on the basis of nuclear track emulsion exposed to secondary beams of the JINR Nuclotron. Contribution of the configuration \mbox{${}^{6}$Be + $\mit{n}$} to the ${}^{7}$Be nucleus structure is \mbox{8 $\pm$ 1\%} which is near the value for the configuration \mbox{${}^{6}$Li + $\mit{p}$}. Distributions over the opening angle of \mbox{$\alpha$-particle} pairs indicate to a simultaneous presence of virtual ${}^{8}$Be$_{g.s.}$ and ${}^{8}$Be$_{2^+}$ states in the ground states of the ${}^{9}$Be and ${}^{10}$C nuclei. The core ${}^{9}$B is manifested in the ${}^{10}$C nucleus with a probability of \mbox{30 $\pm$ 4\%}. Selection of the ${}^{10}$C "white" stars accompanied by ${}^{8}$Be$_{g.s.}$ (${}^{9}$B) leads to appearance in the excitation energy distribution of 2$\alpha$2$\mit{p}$ "quartets" of the distinct peak with a maximum at \mbox{4.1 $\pm$ 0.3 MeV}. ${}^{8}$Be$_{g.s.}$ decays are presented in \mbox{24 $\pm$ 7\%} of 2He + 2H events of the ${}^{11}$C coherent dissociation and \mbox{27 $\pm$ 11\%} of the 3He ones. The channel \mbox{${}^{9}$B + H} amounts \mbox{14 $\pm$ 3\%}. The ${}^{8}$Be$_{g.s.}$ nucleus is manifested in the coherent dissociation \mbox{${}^{10}$B $\to$ 2He + H} with a probability of \mbox{25 $\pm$ 5\%} including \mbox{14 $\pm$ 3\%} of ${}^{9}$B decays. A probability ratio of the mirror channels \mbox{${}^{9}$B + $\mit{n}$} and \mbox{${}^{9}$Be + $\mit{p}$} is estimated to be \mbox{6 $\pm$ 1}.
\end{abstract}


\maketitle

\section*{Introduction}
\begin{figure}[t]
	\centerline{\includegraphics*[width=0.7\linewidth]{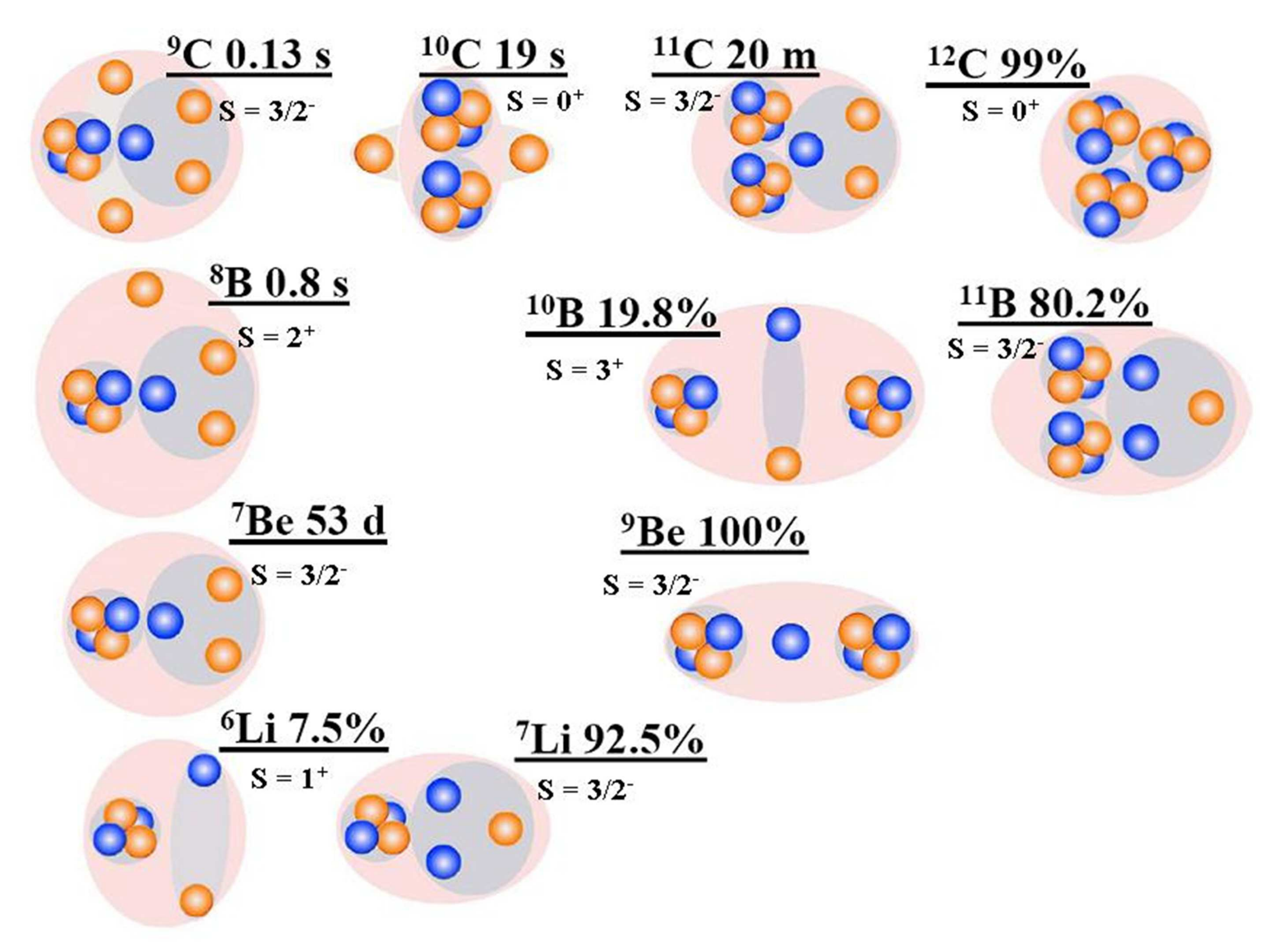}}
	\caption{(Color online). Diagram of cluster degrees of freedom in stable and neutron-deficient nuclei; abundances or lifetimes of isotopes, their spins and parities are indicated; open circles correspond to protons and dark ones -- neutrons; clusters are marked as dark background.}
	\label{fig:Fig.1}
\end{figure}

\noindent The family of nuclei  composing the beginning of the isotope table provides a wholesome "laboratory" which allows one to study the evolution from cluster to shell nuclear structure (Fig. \ref{fig:Fig.1}). As "building blocks" the light nuclei include the lightest clusters having no excited states, namely, \mbox{$\alpha$-particles}, tritons, ${}^{3}$He nuclei, and deuterons, and nucleons which virtual associations coexist in dynamical equilibrium. A pair of \mbox{$\alpha$-particles} can constitute the unstable ${}^{8}$Be nucleus in the ground ${}^{8}$Be$_{g.s.}$ or 1st exited ${}^{8}$Be$_{2^+}$ states. The stable ${}^{7}$Be and ${}^{7}$Li nuclei are important in the structure of the neutron deficient and neutron rich nuclei, respectively. When the nucleons or clusters are added to the ${}^{8}$Be, ${}^{7}$Be and ${}^{7}$Li nuclei the last ones serve as cores in the subsequent stable and radioactive isotopes. Then, the unstable ${}^{9}$B and stable ${}^{9}$Be can possess core roles of an equal importance in heavier nuclear structures. A balanced superposition of the cores, clusters and nucleons in appropriate spin-parity states determine ground state parameters of a corresponding nucleus.

Highlights of nuclear clustering in light nuclei studies obtained recently by means of the nuclear track emulsion (NTE) technique in the framework of the $\href{http://becquerel.jinr.ru/}{BECQUEREL}$ Project \cite{1} are gathered here. In spite of the fact that half a century passed since the NTE development it retains the status of an universal and inexpensive detector \cite{2,3,4}. With an unsurpassed spatial resolution (about 0.5 $\mu$m) NTE of the BR-2 type provides a complete observation of tracks starting from fission fragments and down to relativistic particles. NTE deserve a further use in fundamental and applied research in state-of-art accelerators and reactors, as well as with sources of radioactivity, including natural ones. Application of NTE is especially justified in those pioneering experiments in which nuclear particle tracks can not be reconstructed with the help of electronic detectors. Thus, in the last decade in the framework of the BECQUEREL Project in JINR the NTE technique allowed one to investigate clustering of the nuclei ${}^{7}$Li, ${}^{7,9}$Be, ${}^{8,10}$B, ${}^{9,10}$C and ${}^{12,14}$N in their relativistic dissociation \cite{4,5,6,7,8,9,10,11,12,13}.

However, the production of NTE pellicles which lasted in Moscow for four decades was ended more than ten years ago. The interest in a further application stimulated its reproduction in the MICRON workshop that is part of the company "Slavich" (Pereslavl Zalessky) \cite{13}. At present NTE samples are produced by layers of thickness of 50 to 200 $\mu$m on glass substrates. Supportless pellicles of thickness of the order of 500 $\mu$m are expected to be available soon. Verification of novel NTE in exposures to relativistic particles confirmed that it is similar to the BR-2 one. 

The NTE technique is based on intelligence, vision and performance of researchers using traditional microscopes. Despite widespread interest, its labor consumption causes limited samplings of hundreds of measured tracks which present as a rule only tiny fractions of the available statistics. Implementation of computerized and fully automated microscopes in the NTE analysis allows one to bridge this gap. These are complicated and expensive devices of collective or even remote usage allow one to describe unprecedented statistics of short nuclear tracks. To make such a development purposeful it is necessary to focus on such a topical issue of nuclear physics the solution of which can be reduced to simple tasks of recognition and measurement of tracks in NTE to be solved with the aid of already developed programs. 

Keeping in mind such a perspective competitiveness of NTE in measurements of short \mbox{$\alpha$-particle} and heavy ion tracks on most precise optical microscope KSM with a 90$\times$ objective is demonstrated in series of low energy applications. When measuring decays of ${}^{8}$He nuclei implanted in NTE the possibilities of \mbox{$\alpha$-spectrometry} were verified and the effect of the ${}^{8}$He atom drift was established \cite{14,15,16}. Correlation of \mbox{$\alpha$-particle} triples were studied in disintegrations  of carbon nuclei of NTE composition by 14.1 MeV neutrons \cite{17}. The angular correlations of ${}^{7}$Li and ${}^{4}$He nuclei produced in disintegrations of boron nuclei by thermal neutrons were studied in boron enriched NTE \cite{18}. In this series of exposures the angular resolution of NTE was confirmed to be perfect by expected physical effects which are manifested in the distributions of the opening angles distributions of the products of the studied reactions. 

At CERN, a NTE sample was exposed to 160 GeV muons \cite{18}. The NTE irradiation with these particles makes it possible to study the multifragmentation of nuclei under the effect of a purely electromagnetic probe. Multiphoton exchanges or transitions of virtual photons to mesons may serve as a fragmentation mechanism. The nuclear diffraction mechanism rather than the soft electromagnetic mechanism manifests itself for carbon nuclei splittings to \mbox{$\alpha$-particle} triples. The corroboration of this conclusion is of importance for interpreting not only multifragmentation under the effect of ultrarelativistic muons. It may also serve as a basis for interpreting the multifragmentation of relativistic nuclei in peripheral interactions not leading to the formation of target fragments ("white" stars).

One of the suggested problems is a search for the possibility of a collinear cluster tri-partition \cite{19}. The existence of this phenomenon could be established in observations of such a type of ternary fission of heavy nuclei in which a lightest fragment is emitted in the direction of one of the heavy fragments. Despite distinct observability of fission fragments they can not be fully identified in NTE. However, NTE is valuable due to the combination of the best angular resolution and maximum sensitivity. Besides, it is possible to measure the lengths and thicknesses of tracks, and, thus, to classify the fragments. As an initial stage, to provide statistics of ternary fissions it is suggested to analyze a sufficient NTE area exposed to ${}^{252}$Cf source with an appropriate density of tracks of \mbox{$\alpha$-particles} and spontaneous fission fragments \cite{20}. Such an approach will be developed by a NTE with an admixture of the ${}^{252}$Cf isotope \cite{5,6}. Another option is exposure by thermal neutrons of NTE manufactured with a ${}^{235}$U isotope addition.

On high energy side discussed below novel samples of NTE were exposed quite recently to the secondary beam of relativistic nuclei ${}^{11}$C of the JINR Nuclotron which allowed one to include a clustering of the ${}^{11}$C nucleus into the general pattern already relying on data on the light nuclei including radioactive ones. The ${}^{11}$C data stimulated an additional analysis of previous ${}^{10}$C and ${}^{10}$B exposures.

\section*{Nuclear clustering in relativistic dissociation}

\noindent Consideration of the nucleosynthesis chains toward ${}^{10,11}$B, ${}^{10,11}$C and ${}^{12}$N via the "hot breakout" ${}^{7}$Be(${}^{3}$He,$\gamma$)${}^{10}$C(e$^+$,$\nu$)${}^{10}$B assists to recognize relations between their structures and, in particular, importance of the unbound nuclei in them. The ${}^{10}$C synthesis processing due to an increase of \mbox{$\alpha$-clustering} provides an energy "window" for the formation of intermediate states with unstable nuclei \mbox{${}^{9}$B + $\mit{p}$}, \mbox{${}^{8}$Be$_{2^+}$ + 2$\mit{p}$} and \mbox{${}^{6}$Be + $\alpha$}. These clusters are preserved in subsequent reactions \mbox{${}^{10}$C(e$^+$,$\nu$)${}^{10}$B($\mit{p}$,$\gamma$)${}^{11}$C(e$^+$,$\nu$)${}^{11}$B}. The "window" of the reaction \mbox{${}^{7}$Be(${}^{4}$He,$\gamma$)${}^{11}$C} allows only an association of the ${}^{7}$Be and ${}^{4}$He clusters, also contributing to the ${}^{11}$C and ${}^{10}$B structure. Thus, a hidden variety of the virtual configurations in the nuclei ${}^{10,11}$C and ${}^{10,11}$B can be populated via electromagnetic transitions from the real ones. In turn, these nuclei provide a basis for capture reactions of protons or the He isotopes (or in neutron exchange) for synthesis of the subsequent nuclei which leads to a translation of the preceding structures. Obviously, the unstable nuclei ${}^{8}$Be and ${}^{9}$B play a key role in a general pattern of nuclear clustering and despite complicated observability their contribution deserves to be studied in detail over an available variety of light nuclei and physical mechanisms.

The cluster structure of light nuclei including radioactive ones in relativistic-fragmentation processes is a central topic of the BECQUEREL project which continues the tradition of use of the NTE technique \cite{5}. Such reactions are under study by means of NTE stacks longwise exposed to primary and secondary beams of relativistic nuclei of the JINR Nuclotron. Among the events of fragmentation of relativistic nuclei, those of their coherent dissociation to narrow jets of fragments are especially valuable for studying nuclear clustering. Coherent dissociation does not feature either slow fragments of NTE composing nuclei or charged mesons. This empirical feature allows one to assume a glancing character of such collisions and that excitations of relativistic nuclei under study are minimal. A main underlying mechanism of coherent dissociation is nuclear diffraction interaction processing without nuclear density overlap and angular momentum transfer. The experimental method in question has already furnished unique information compiled in \cite{5} about cluster aspects of the structure of the whole family of light nuclei, including radioactive ones. 

\begin{figure}[t]
	\centerline{\includegraphics*[width=1\linewidth]{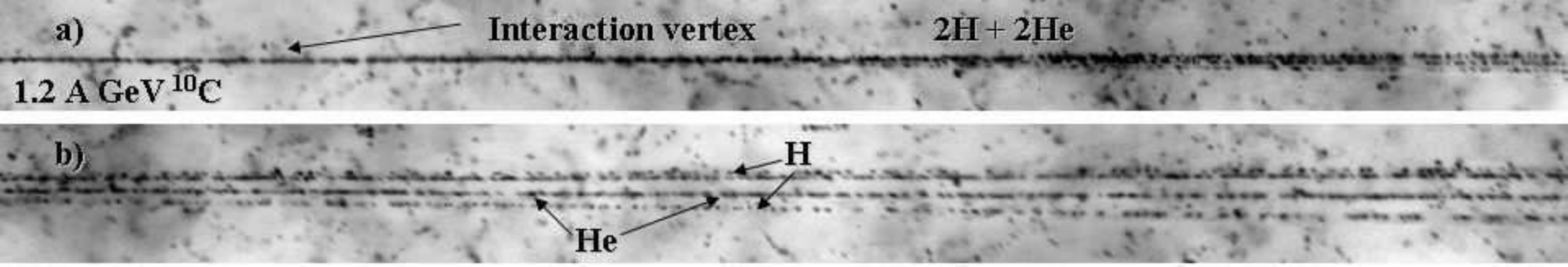}}
	\caption{Macrophoto of the coherent dissociation event of 1.2 $\mit{A}$ GeV ${}^{10}$C nucleus into pairs of He and H nuclei; a) primary track, approximate position of interaction vertex and appearance of fragment tracks and b) tracks of fragments are resolved; opening angles between tracks are \mbox{$\Theta_{2He}$= 5.9 mrad}, \mbox{$\Theta_{HeH}$ = 8.6,} 16.6, 3.0, 17.6 mrad, \mbox{$\Theta_{2p}$ = 20.1 mrad}. Both 2HeH triples in the event correspond to ${}^{9}$B decays.}
	\label{fig:Fig.2}
\end{figure}

Events of coherent dissociation are called "white" stars because of the absence of tracks of strongly ionizing particles (Fig. \ref{fig:Fig.2}). The term "white" star reflects aptly a sharp "breakdown" of the ionization density at the interaction vertex upon going over from the primary-nucleus track to secondary tracks within a $6^\circ$ cone at 1.2 $\mit{A}$ GeV. This special feature generates a fundamental problem for electronic methods because more difficulties should be overcome in detecting events where the degree of dissociation is higher. On the contrary, such events in NTE are observed and interpreted in the most straightforward way, and their distribution among interaction channels characterized by different compositions of charged fragments is determined exhaustively. 

The probability distribution of the final configurations of fragments in "white" stars makes it possible to reveal their contributions to the structure of nuclei under consideration. We assumed that, in the case of dissociation, specific configurations arise at random without sampling and that the dissociation mechanism itself does not lead to the sampling of such states via angular-momentum or isospin exchange. By and large, available results confirm the assumption that the cluster features of light nuclei determine the pattern of their relativistic dissociation. At the same time, events that involve the dissociation of deeply bound cluster states and which cannot arise at low collision energy are detected.

\begin{table}[t]
	\caption{Distribution ${}^{7}$Be "white" stars over charge channels.}
	\begin{center}\begin{tabular}{c c} \hline 
			Channel & ${}^{7}$Be \cite{8} \\ \hline
			2He &	115 (40 \%) \\
			He + 2H &	157 (54 \%) \\
			Li + H &	14 (5 \%) \\
			4H &	3 (1 \%)\ \\ \hline 
			\label{tabular:Tab.1}
		\end{tabular}
	\end{center}
\end{table}

\begin{table}[t]
	\caption{Distribution ${}^{10}$B and ${}^{8}$B "white" stars over charge channels.}
	\begin{center}\begin{tabular}{c c c} \hline 
			Channel & ${}^{10}$B  & ${}^{8}$B \\ \hline
			Be + H	& 2 (1 \%) &	25 (48 \%) \\
			2He + H	& 108 (78 \%) &	14 (27 \%) \\
			He + 3H	& 18 (13 \%)	& 12 (23 \%) \\
			Li + He	& 5 (4 \%)  &	- \\
			Li + 2H &	5 (4 \%) & 	- \\
			5H &	2 & 	- \\ \hline
			\label{tabular:Tab.2} 
		\end{tabular}
	\end{center}
\end{table}

Data on the previously studied nuclei are valuable ingredients of the ongoing analysis, and deserve a brief description. The feature of ${}^{7}$Be dissociation is an approximate equality of probability of the main channels 2He and \mbox{He + 2H} of coherent dissociation (Table \ref{tabular:Tab.1}). Their ratio is equal to \mbox{0.7 $\pm$ 0.1} \cite{11}. Recently obtained statistics of 140 ${}^{10}$B "white" stars is presented in Table \ref{tabular:Tab.2}. The channel \mbox{2He + H} leads reaching about 77\%. Events of channel \mbox{He + 3H} are 13\%. 4\% of the events contain both fragments Li and He. Not more than 2\% of the events contain fragments Be and H, indicating the minor probability of configuration \mbox{${}^{9}$Be + $\mit{p}$} in the ${}^{10}$B structure.

In contrast, the contribution of the channel \mbox{${}^{7}$Be + $\mit{p}$} in the ${}^{8}$B coherent dissociation is a leading one while configurations containing only clusters of He and H is estimated at 50\%. One can observe approximately identical fractions of the \mbox{2He + H} and \mbox{He + 2H} channels, and if one of H is subtracted this fact is compatible with the dissociation of ${}^{7}$Be as the core of the ${}^{8}$B nucleus.

\section*{The unstable nuclei in dissociation of the ${}^{9}$Be, ${}^{10}$C and ${}^{7}$Be nuclei}

\noindent Reconstruction of the decays of relativistic ${}^{8}$Be and ${}^{9}$B nuclei is possible by the energy variable \mbox{$\mit{Q}$ = $\mit{M}^*$ - $\mit{M}$}, where \mbox{$\mit{M}^{*2}$ = $\sum$(P$_i$$\cdot$P$_k$)}, $\mit{M}$ is the total mass of fragments, and P$_{i,k}$ are their 4-momenta defined under the assumption of conservation of an initial momentum per nucleon by fragments. When the identification of relativistic fragment can be reasonably supposed the quasi-invariant variable $\mit{Q}$ allows one to estimate the excitation energy of their complex ensembles uniting all angular measurements in an event. For the "white" stars of ${}^{9}$Be and ${}^{10}$C nuclei the assumption that He fragments correspond to ${}^{4}$He nuclei ($\alpha$), and H ones in ${}^{10}$C -- ${}^{1}$H ($\mit{p}$) is justified. Then ${}^{8}$Be and ${}^{9}$B identification is reduced to measurements of the opening angles between the directions of fragment emission. The experimental details and development of these investigations and their illustrations are presented in \cite{5}. 

\begin{figure}[t]
	\centerline{\includegraphics*[width=0.4\linewidth]{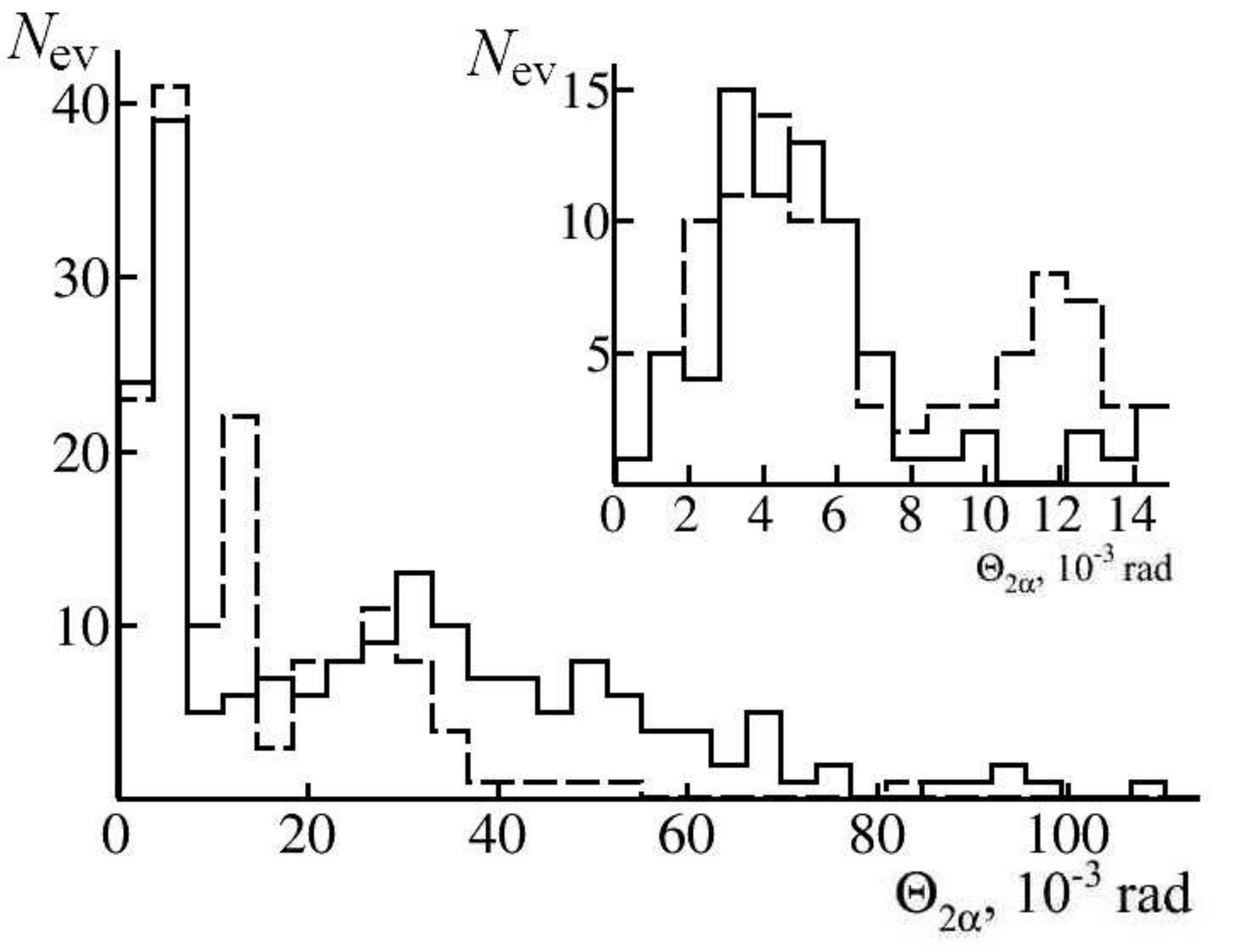}}
	\caption{Distributions over the opening angle $\Theta_{2He}$ of $\alpha$-particle pairs in "white" stars \mbox{${}^{10}$C $\to$ 2He + 2H} (solid) and \mbox{${}^{9}$Be $\to$ 2He} (dashed); on the insertions: the $\Theta_{2He}$ distribution in interval $\Theta_n$.}
	\label{fig:Fig.3}
\end{figure}

Distributions over the opening angle $\Theta$$_{2He}$ for pairs of He fragments of "white" stars \mbox{${}^{9}$Be $\to$ 2He} and \mbox{${}^{10}$C $\to$ 2He + 2H} (82\% of the ${}^{10}$C statistics) produced at energy of 1.2 $\mit{A}$ GeV are presented in Fig. \ref{fig:Fig.3}. In both cases the values of $\Theta$$_{2He}$ of \mbox{75 - 80\%} of the pairs are distributed about equally in the intervals of \mbox{0 $<$ $\Theta$$_{n(arrow)}$ $<$ 10.5 mrad} and \mbox{15.0$<\Theta_{w(ide)}<$ 45.0 mrad}. The remaining pairs are attributed to intervals \mbox{10.5 $<$ $\Theta$$_{m(edium)}$ $<$ 15.0 mrad} and "widest" of \mbox{15.0$< \Theta_{v(ery)w(ide)} <$ 45.0 mrad}. The distribution over the $\mit{Q}$ variable is directly correlated with the $\Theta_{2He}$ one. It is pointing out that "narrow" pairs of $\Theta_n$ are produced via ${}^{8}$Be$_{g.s.}$ while pairs $\Theta_w$ via ${}^{8}$Be$_{2^+}$. Besides, for the ${}^{9}$Be case there is a peak in the interval $\Theta_m$ reflecting its level 5/2$^-$ \mbox{(2.43 MeV)}. Fractions of events in the intervals $\Theta_n$ and $\Theta_w$ are equal to \mbox{0.56 $\pm$ 0.04} and \mbox{0.44 $\pm$ 0.04} for ${}^{9}$Be, while for ${}^{10}$C \mbox{0.49 $\pm$ 0.06} and \mbox{0.51 $\pm$ 0.06}, i. e. they practically coincide. They indicate to a simultaneous presence of virtual ${}^{8}$Be$_{g.s.}$ and ${}^{8}$Be$_{2^+}$ states in the ground states of the ${}^{9}$Be and ${}^{10}$C nuclei.  Elongation above \mbox{40 mrad} of the ${}^{10}$C  $\Theta$$_{2He}$ distribution can be due to the channel \mbox{${}^{4}$He + ${}^{6}$Be}.

\begin{figure}[t]
	\centerline{\includegraphics*[width=0.7\linewidth]{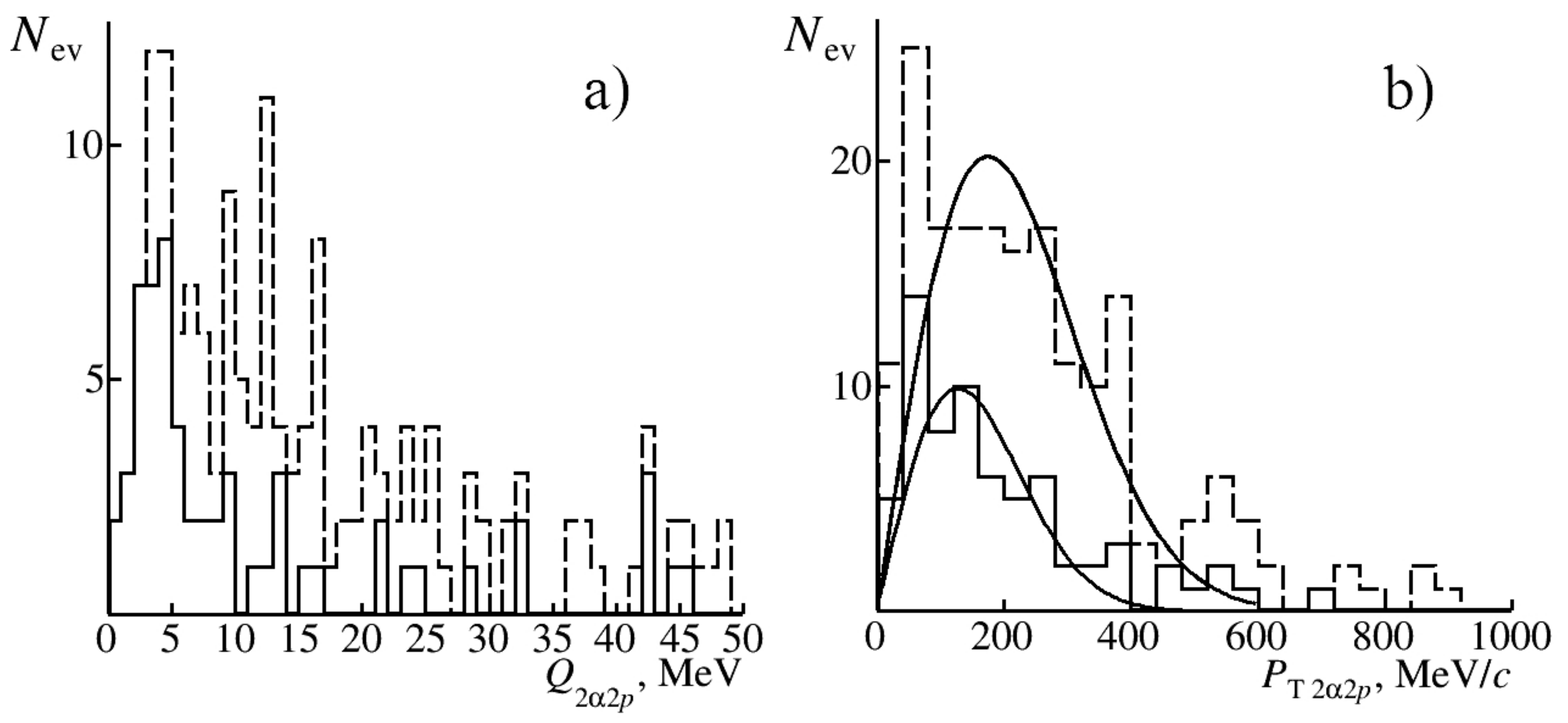}}
	\caption{Distributions over energy $\mit{Q}_{2\alpha 2p}$ (a) and total transverse momentum $\mit{P}_{T2\alpha 2p}$ (b) of all "white" stars \mbox{${}^{10}$C $\to$ 2He + 2H} (dashed) and the ones with the presence of ${}^{9}$B (solid).}
	\label{fig:Fig.4}
\end{figure}

Earlier, basing on the $\mit{Q}_{2\alpha p}$ energy distribution of the triples \mbox{2$\alpha$ + $\mit{p}$} from the "white" stars \mbox{${}^{10}$C $\to$ 2$\alpha$ + 2$\mit{p}$} it is concluded that in the ${}^{10}$C nucleus the core ${}^{9}$B is manifested with a probability of \mbox{30 $\pm$ 4\%}, and the ${}^{8}$Be$_{g.s.}$ decays are arise always through the ${}^{9}$B decays. An interesting feature is manifested in excitation energy distribution $\mit{Q}_{2\alpha 2p}$ of 2$\alpha$2$\mit{p}$ "quartets" provided by completeness of their observation. Fig. \ref{fig:Fig.4} a shows distribution $\mit{Q}_{2\alpha 2p}$ of all "white" stars \mbox{${}^{10}$C $\to$ 2He + 2H} which appears at a first glance to be scattered. The distribution $\mit{Q}_{2\alpha 2p}$ of the ${}^{10}$C stars containing ${}^{9}$B decays features the distinct peak with a maximum at \mbox{4.1 $\pm$ 0.3} MeV at RMS of 2.0 MeV. Its width is determined by the accepted momentum approximation. The peak statistics present \mbox{17 $\pm$ 4\%} of the total number of the ${}^{10}$C "white" stars or \mbox{65 $\pm$ 14\%} of those containing ${}^{9}$B decays.

\begin{figure}[t]
	\centerline{\includegraphics*[width=0.5\linewidth]{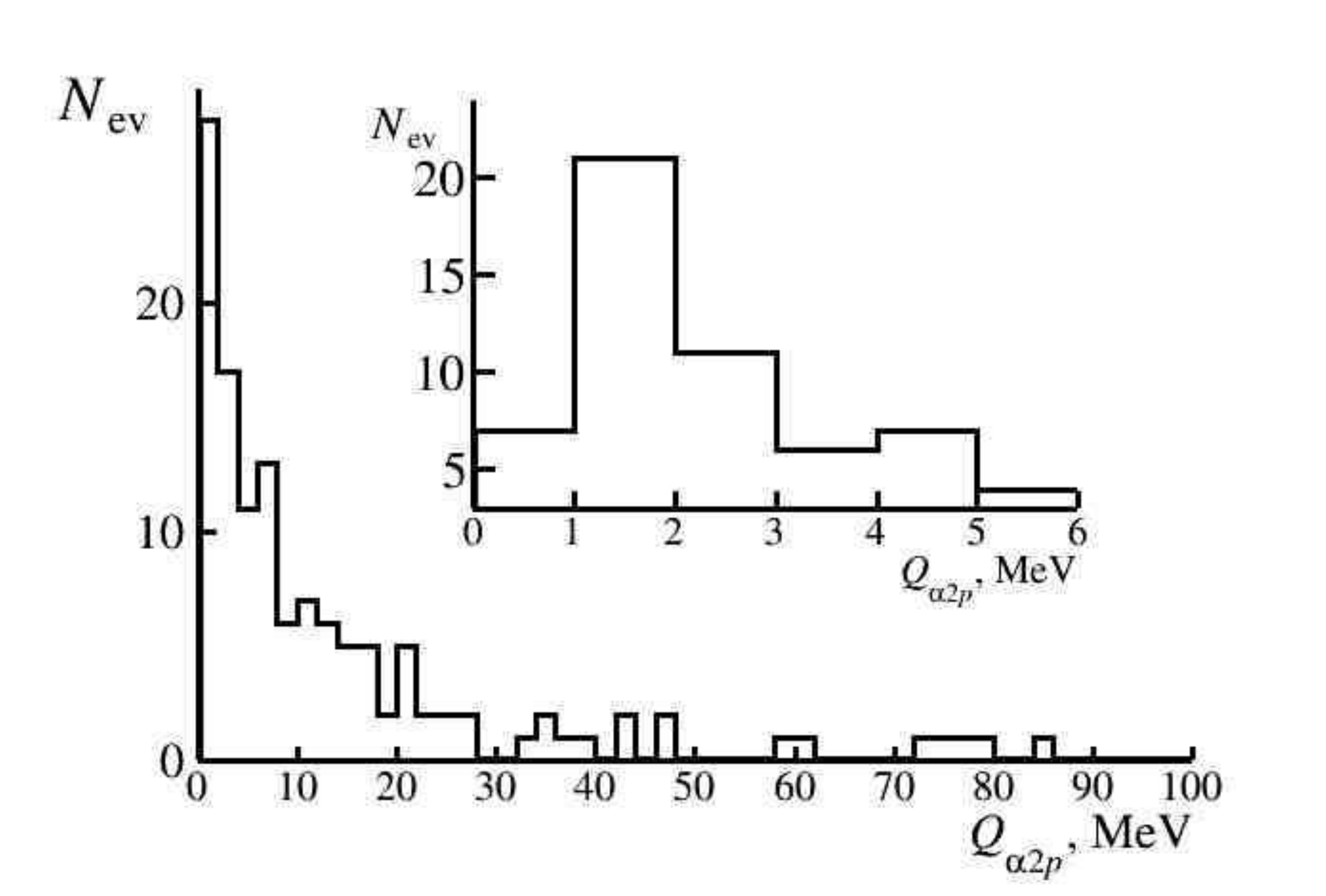}}
	\caption{Distributions over energy $\mit{Q}_{\alpha 2p}$ in measured "white" stars \mbox{${}^{7}$Be $\to$ He + 2H}; on insertion: the enlarged $\mit{Q}_{\alpha 2p}$ distribution in the ${}^{6}$Be region.}
	\label{fig:Fig.5}
\end{figure}

Distribution over a total momentum $\mit{P}_{T2\alpha 2p}$ of all 2$\alpha$2$\mit{p}$ ensembles (Fig. \ref{fig:Fig.4}b) is described by a Rayleigh function with the parameter \mbox{$\sigma$ = 175 $\pm$ 10 MeV/c} while in the case of the presence of ${}^{8}$Be$_{g.s.}$ (${}^{9}$B) decays it is significantly less \mbox{$\sigma$ = 127 $\pm$ 16 MeV/c}. Not competing in statistics and resolution \cite{4} our observation of such a state manifesting in extra narrow 2$\alpha$2$\mit{p}$ jets is grounded on selection of evidently glancing collisions which reduce dramatically a continuum contribution. It is worth noting the observation of a single "white" star 2$\alpha$2$\mit{p}$ having $\mit{Q}_{2\alpha 2p}$ equal to \mbox{0.77 MeV} in which both 2$\alpha$$\mit{p}$ triples correspond to ${}^{9}$B decays with $\mit{Q}_{2\alpha p}$ of 0.22 and \mbox{0.67 MeV}, $\mit{Q}_{2\alpha}$ of \mbox{0.14 MeV} and $\mit{Q}_{\alpha 2p}$ of 0.64 and \mbox{0.6 MeV} (Fig. \ref{fig:Fig.5}).

\begin{figure}[t]
	\centerline{\includegraphics*[width=0.8\linewidth]{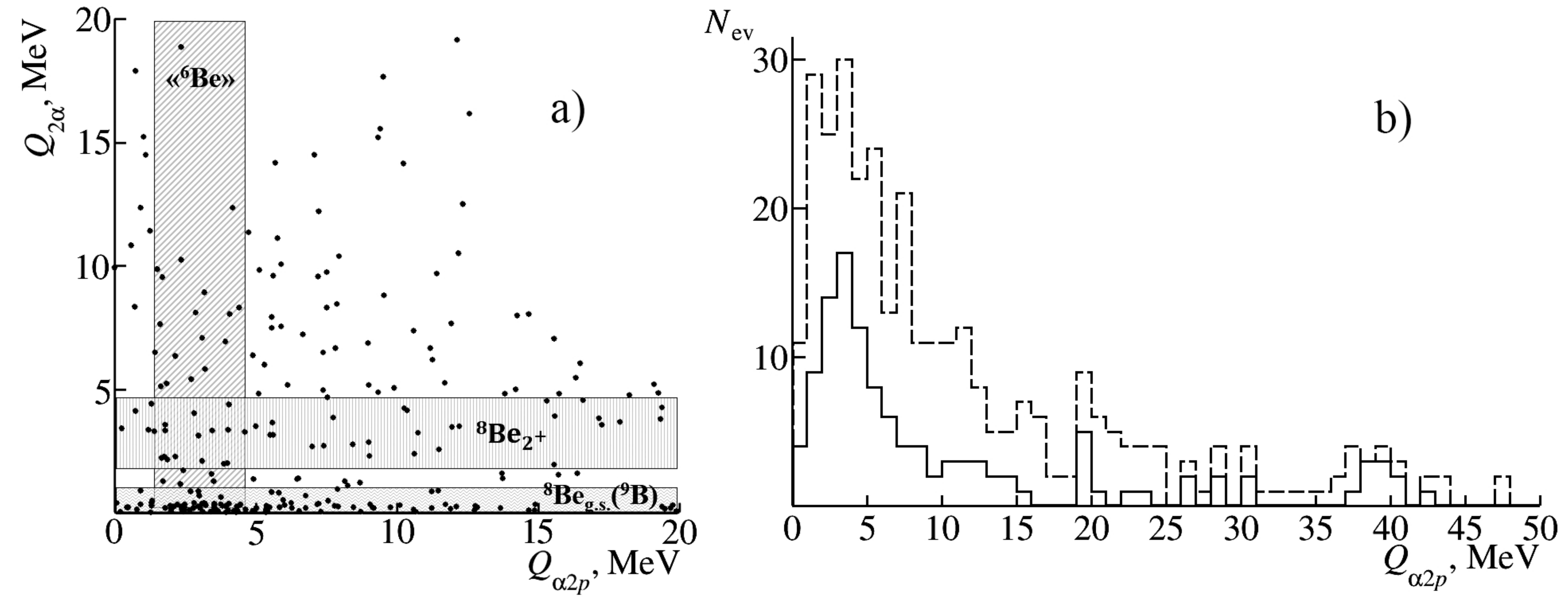}}
	\caption{Distributions of "white" stars \mbox{${}^{10}$C $\to$ 2He + 2H} over energy $\mit{Q}_{2\alpha}$ and $\mit{Q}_{\alpha 2p}$ (a, expected regions of decays ${}^{6}$Be, ${}^{8}$Be$_{g.s.}$ (${}^{9}$B) and ${}^{8}$Be$_{2^+}$ are shown). Distributions of $\alpha$2$\mit{p}$ triples over energy $\mit{Q}_{\alpha 2p}$ in all "white" stars \mbox{${}^{10}$C $\to$ 2He + 2H} (b, dashed histogram) and with presence of ${}^{8}$Be$_{g.s.}$ (${}^{9}$B) decays in them (b, solid histogram).}
	\label{fig:Fig.6}
\end{figure}

High statistics of "white" stars \mbox{He + 2H} produced by 1.2 $\mit{A}$ GeV ${}^{7}$Be nuclei \cite{5} allows one to estimate the contribution of the unbound ${}^{6}$Be nucleus in the distribution over $\mit{Q}_{\alpha 2p}$ (Fig. \ref{fig:Fig.5}). \mbox{27 $\pm$ 5\%} of 130 events in the channel \mbox{He + 2H} can be attributed to ${}^{6}$Be decays by the condition \mbox{$\mit{Q}_{\alpha 2p}$ $<$ 6 MeV}. Contribution of the configuration \mbox{${}^{6}$Be + $\mit{n}$} to the ${}^{7}$Be structure is estimated at a level of \mbox{8 $\pm$ 1\%} which is near of the value of \mbox{5 $\pm$ 1\%} for the configuration \mbox{${}^{6}$Li + $\mit{p}$}. Detection efficiency of the last one is somewhat less. So, the difference of probabilities can be considered as unsignificant.

Determination of the ${}^{6}$Be decay region gives an opportunity to estimate a possible ${}^{6}$Be contribution to the ${}^{10}$C "white" stars. Fig. \ref{fig:Fig.6}a shows correlation between $\mit{Q}_{2\alpha}$ and $\mit{Q}_{\alpha 2p}$. A peak at \mbox{3-4 MeV} in the total distribution $\mit{Q}_{\alpha 2p}$ covers totally the expected ${}^{6}$Be signal. The peak is becoming profound when ${}^{8}$Be$_{g.s.}$ decays in the stars are demanded (Fig. \ref{fig:Fig.6}b). Surprisingly, but one can not separate the ${}^{6}$Be and ${}^{8}$Be$_{g.s.}$ decays and have to assume that ${}^{6}$Be and ${}^{8}$Be$_{g.s.}$ are produced as interfering parts of 2$\alpha$2$\mit{p}$ ensembles.

\section*{Dissociation of the ${}^{11}$C nucleus}

\noindent The ${}^{11}$C isotope combines peculiarities of stable nuclei with a pronounced $\alpha$-particle clustering and nuclei at the boundary of proton stability where ${}^{3}$He clustering is of the same importance. Interactions in the ensemble \mbox{2${}^{4}$He + ${}^{3}$He} lead to formation of the weakly bound configurations \mbox{${}^{7}$Be + $\alpha$} \mbox{(7.6 MeV)}, \mbox{${}^{10}$B + $\mit{p}$} \mbox{(8.7 MeV)} and \mbox{${}^{3}$He + ${}^{8}$Be} (9.2 MeV).

Exposure of a series of NTE samples is performed in the secondary beam of relativistic nuclei ${}^{11}$C of the JINR Nuclotron. Nuclei ${}^{11}$C were produced in fragmentation of nuclei ${}^{12}$C of energy 1.2 $\mit{A}$ GeV. Reduced thickness and glass substrates of an experimental batch of NTE are factors which do not allow carrying out an analysis with scanning along beam tracks without sampling. Therefore, scanning of the NTE layer was carried out on transverse strips. Tracks corresponding doubly and singly charged relativistic particles are determined visually. Dominance in the beam of C nuclei allows specifying charges of heavier fragments in "white" stars as values missing up to six charge units.

To date, 144 "white" stars with a total charge of relativistic fragments equal to 6 are found in scanned layers. Their distribution over charge states is presented jointly with comparable data on the isotopes ${}^{10}$C and ${}^{9}$C in Table \ref{tabular:Tab.3}. Table \ref{tabular:Tab.3} demonstrates the specific nature of "white" stars of each of the isotopes and compliance of the performed exposures to the mass numbers of the C isotopes. In the study of coherent dissociation of relativistic ${}^{12}$C nuclei all found 100 "white" stars emerged in a single channel \mbox{${}^{12}$C $\to$ 3He} clearly reflecting the 3$\alpha$-particle clustering of this nucleus. The key observation then became decays of unbound relativistic \mbox{${}^{8}$Be nuclei which gave the contribution of not less than 20\%.}

Events containing only the relativistic isotopes of He and H (77\%), in particular, \mbox{2He + 2H} dominate among the ${}^{11}$C "white" stars. The ratio of this channel statistics to the statistics of the channel \mbox{He + 4H} is \mbox{5 $\pm$ 2}. It does not correspond to only dissociation of the ${}^{7}$Be core mentioned above. In contrast to the previously studied neutron deficient nuclei the significant share of events \mbox{Li + He + H} is observed, which could correspond to \mbox{${}^{6}$Li + ${}^{4}$He + $\mit{p}$}. There are no events \mbox{Be + 2H} which could correspond to \mbox{${}^{9}$Be + 2$\mit{p}$}. However, there is a significant fraction of events \mbox{Be + He} in which the isotope ${}^{7}$Be is uniquely determined. Most likely that the 3He channel corresponds to the configuration 2${}^{4}$He${}^{3}$He which can arise as from the break-ups of the core nuclei ${}^{8}$Be and ${}^{7}$Be and "dilute" 3He states. Additional contribution to the multiple He and H channels can be made cluster dissociation ${}^{6}$Li, as a separate element of the ${}^{11}$C according to the bond structure \mbox{$\alpha$ + $\mit{d}$}. Figuratively being expressed the charge topology distributions have an individual character to ${}^{11}$C which different from the other isotopes as a kind of "autograph".

\begin{table}[t]
	\caption{Distribution of "white" stars produced by the C isotopes at 1.2 $\mit{A}$ GeV over charge channels.}
	\begin{center}\begin{tabular}{c c c c} \hline 
			Channel & ${}^{11}$C \cite{12}  & ${}^{10}$C \cite{9} & ${}^{9}$C \cite{5} \\ \hline
			B + H	& 6 (4 \%) & 1 (0.4 \%) & 15 (14 \%) \\
			Be + He & 17 (12 \%) & 6 (2.6 \%) & -	 \\
			Be + 2H & - & - & 16 (15 \%)	 \\
			3He & 26 (18 \%) & 12 (5.3 \%) & 16 (15 \%)	\\
			2He + 2H & 72 (50 \%) & 186 (82 \%) & 24 (23 \%) \\
			He + 4H & 15 (11 \%) & 12 (5.3 \%) & 28 (27 \%) \\ 
			Li + He + H & 5 (3 \%) & - & -\\
			Li + 3He & -& 1 (0.4 \%) & 2 (2 \%) \\ \hline 
			\label{tabular:Tab.3}
		\end{tabular}
	\end{center}
\end{table}
\begin{figure}[t]
	\centerline{\includegraphics*[width=0.75\linewidth]{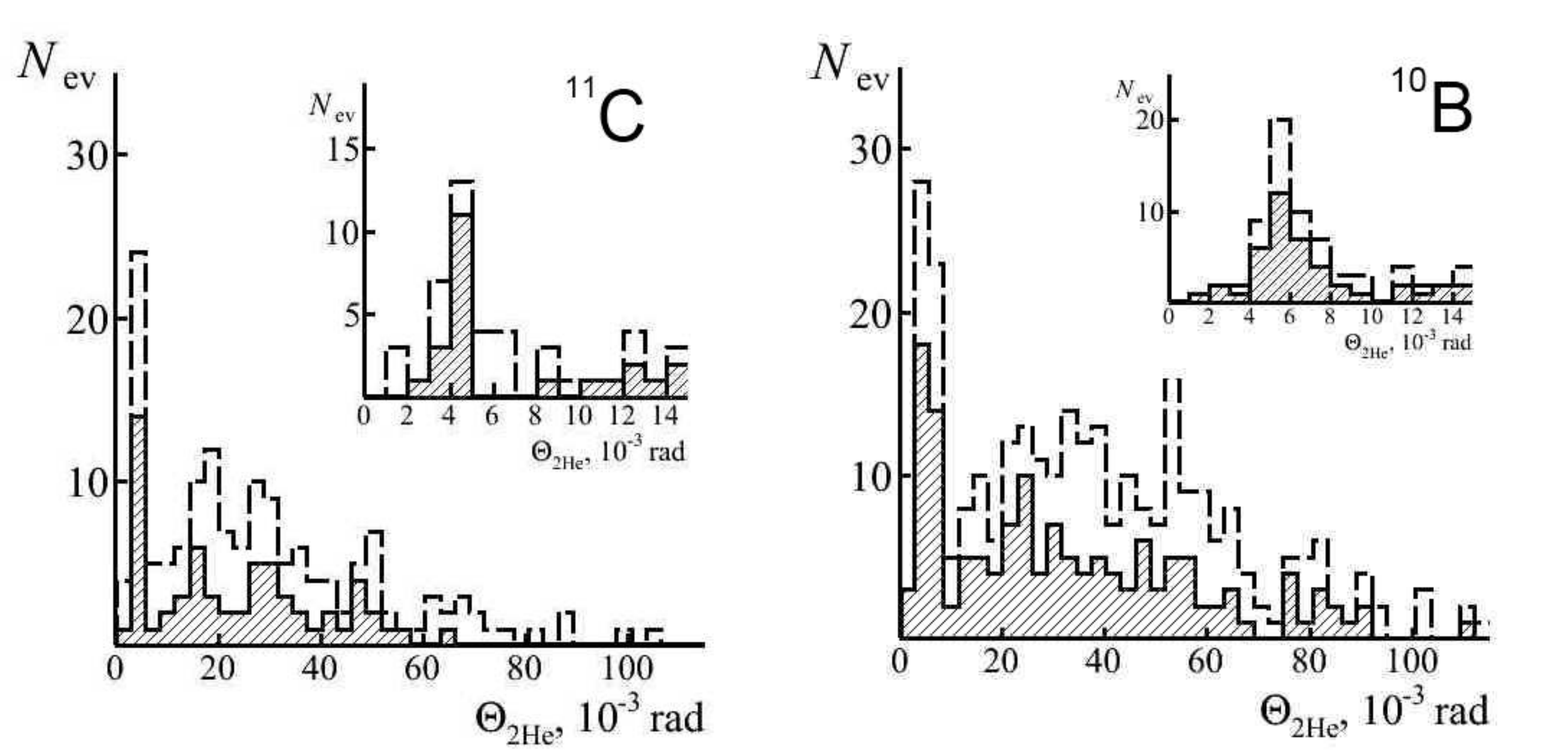}}
	\caption{Distributions over opening angle $\Theta_{He}$ between measured directions of He fragments in all ${}^{11}$C stars \mbox{2He + 2H} (dashed), ${}^{11}$C "white"  stars \mbox{2He + 2H} (hatched), all ${}^{10}$B stars \mbox{2He + H} (dashed) and ${}^{10}$B "white"  stars \mbox{2He + H} (hatched).}
	\label{fig:Fig.7}
\end{figure}

\begin{figure}[t]
	\centerline{\includegraphics*[width=0.4\linewidth]{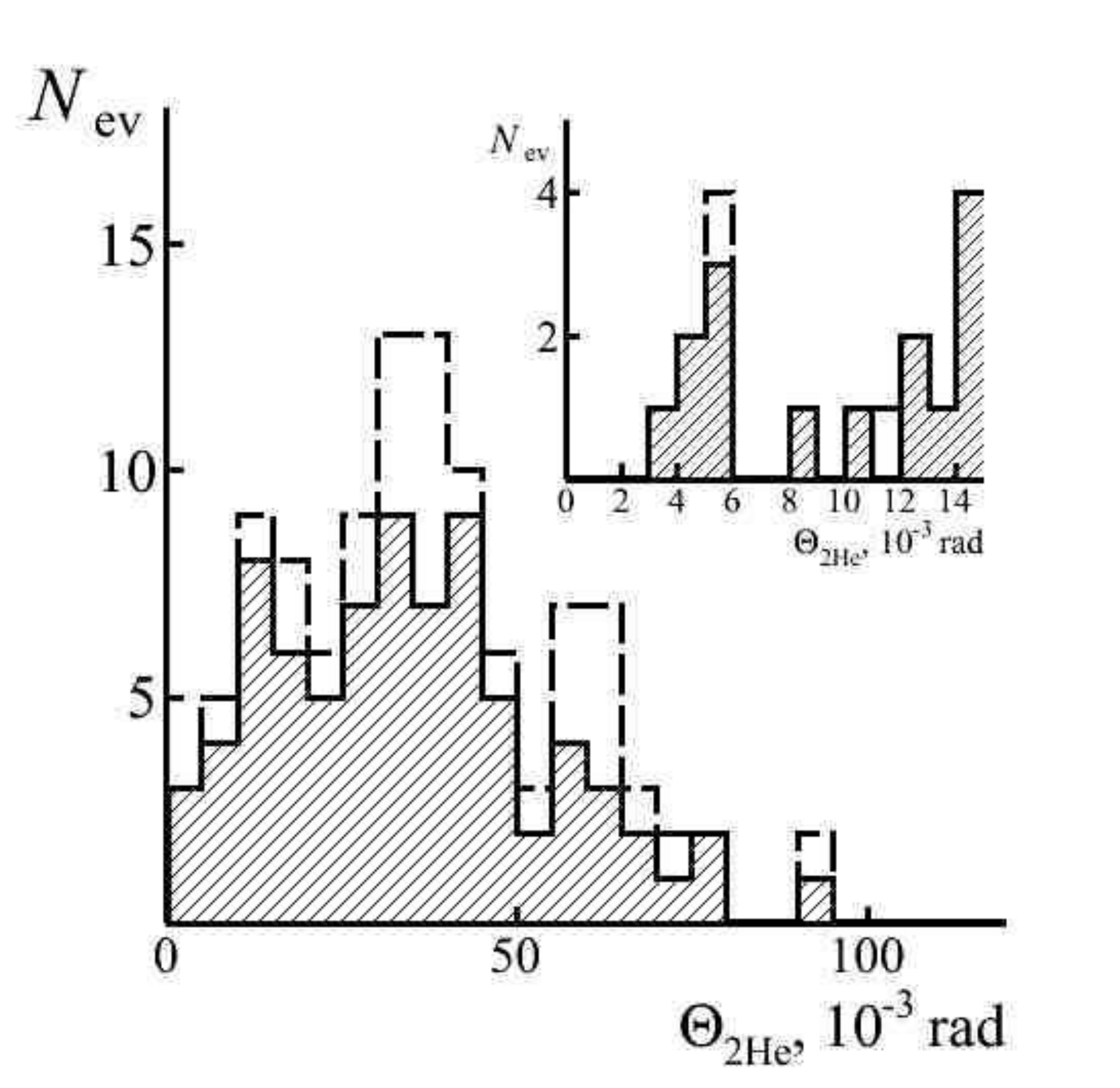}}
	\caption{Distributions over opening angle $\Theta_{He}$ between measured directions of He fragments in all ${}^{11}$C stars 3He (dashed) and ${}^{11}$C "white"  stars 3He (hatched).}
	\label{fig:Fig.8}
\end{figure}

Emission angles of fragments were measured in 156 dissociations ${}^{11}$C $\to$ 2He + 2H among (212 found) including 62 "white" (72 found). The distributions over the opening angle $\Theta_{2He}$ of He fragments (Fig. \ref{fig:Fig.7}, left) points to the presence of 16 decays ${}^{8}$Be$_{g.s.}$ in "white" stars amounting them \mbox{24 $\pm$ 7\%} in this channel. In the same way, 26 "white" stars of the 3He channel contain 7 decays ${}^{8}$Be$_{g.s.}$ (Fig. \ref{fig:Fig.8}) amounting \mbox{27 $\pm$ 11\%} in this channel and \mbox{5 $\pm$ 2\%} of the channel \mbox{${}^{8}$Be$_{g.s.}$ + ${}^{3}$He} in the overall statistics (Table \ref{tabular:Tab.3}). Besides, the distributions allow one to assume a strong contribution of ${}^{8}$Be$_{2^+}$ decays but it is a subject of future detailed consideration.

\begin{figure}[t]
	\centerline{\includegraphics*[width=0.75\linewidth]{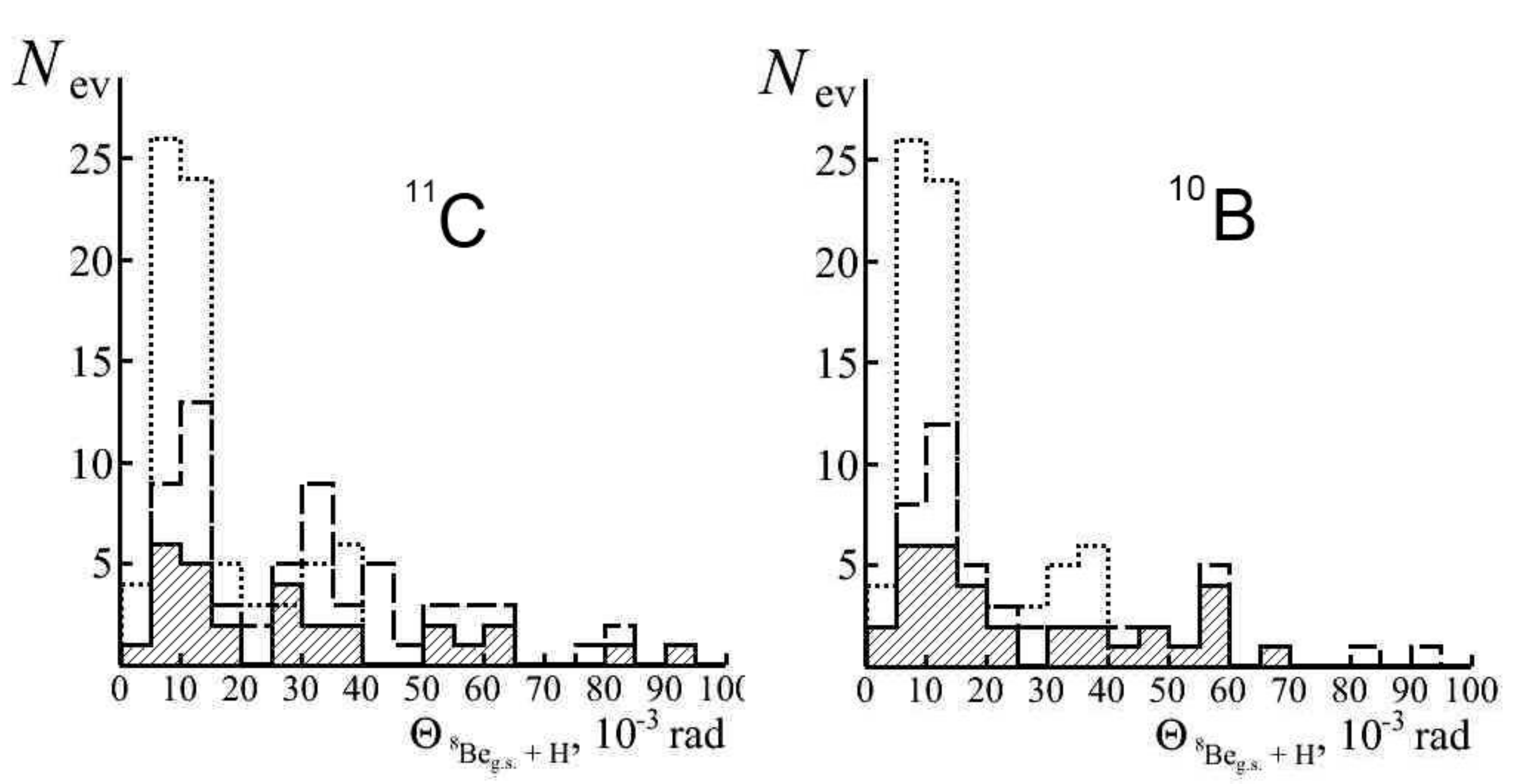}}
	\caption{Distributions over opening angle $\Theta_{{}^{8}Be_{g.s.} + H}$ between measured directions of fragments ${}^{8}$Be$_{g.s.}$ and H fragments in ${}^{10}$C "white" stars  (dotted), all ${}^{11}$C stars  (left, dashed), ${}^{11}$C "white"  stars (left, hatched), all ${}^{10}$B stars  (right, dashed) and ${}^{10}$B "white"  stars (right, hatched).}
	\label{fig:Fig.9}
\end{figure}

The virtual ${}^{9}$B nucleus can exist in the ${}^{11}$C nucleus as an independent component or as a component of a virtual core ${}^{10}$B or ${}^{10}$C. ${}^{9}$B decays are identified by the small opening angle between directions of ${}^{8}$Be$_{g.s.}$ and each one of H fragments \mbox{$\Theta_{{}^{8}Be_{g.s.} + H}$ $<$ 25 mrad} (Fig. \ref{fig:Fig.9}). In the same way, 14 ${}^{9}$B decays are identified in "white" stars \mbox{${}^{11}$C $\to$ 2He + 2H} (Fig. \ref{fig:Fig.9}, left). Important conclusion is that being almost the same as the ${}^{8}$Be$_{g.s.}$ number the ${}^{9}$B decay number points on predominantly cascade production ${}^{8}$Be$_{g.s.}$ via ${}^{9}$B like in the ${}^{10}$C case. On this ground the channel \mbox{${}^{9}$B + H amounts 14 $\pm$ 3\% of the channel  in the ${}^{11}$C "white" star statistics (Table \ref{tabular:Tab.3})}.

\begin{figure}[t]
	\centerline{\includegraphics*[width=0.4\linewidth]{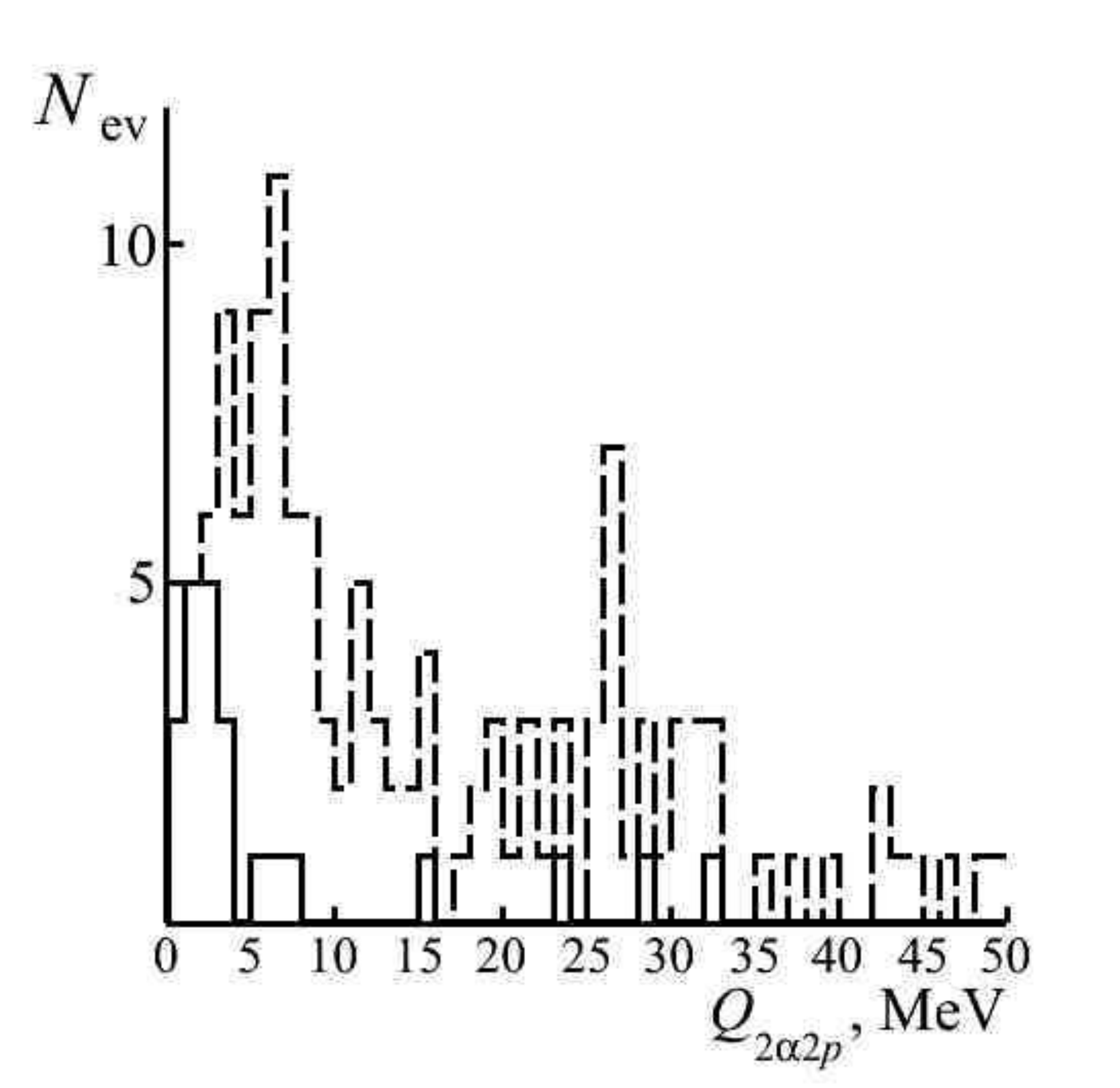}}
	\caption{Distributions over energy $\mit{Q}_{2\alpha 2p}$  of all found stars \mbox{${}^{11}$C $\to$ 2He + 2H} (dashed) and the ones with the presence of ${}^{9}$B (solid).}
	\label{fig:Fig.10}
\end{figure}

Preliminary, correspondence of H to p and He to $\alpha$ can be assumed in calculation $\mit{Q}_{2\alpha 2p}$. Worth mentioning is the lowest energy peak in the distribution $\mit{Q}_{2\alpha 2p}$ of 18 found stars \mbox{${}^{11}$C $\to$ 2He + 2H} containing ${}^{9}$B decays (Fig. \ref{fig:Fig.10}). In two cases both 2$\alpha$$\mit{p}$ triples correspond to ${}^{9}$B decays. Having the same meaning as one in the ${}^{11}$C case it is characterized by a somewhat less mean value of \mbox{2.7 $\pm$ 0.4 MeV} at RMS of \mbox{2.0 MeV}. A tendency can be noted that the ${}^{9}$B condition selects "coldest" events among stars \mbox{2He + 2H}.

\section*{Resumed analysis of dissociation of the ${}^{10}$B nucleus}

\noindent The early analysis of the NTE exposured in 2001 to 1 $\mit{A}$ GeV ${}^{10}$B nuclei has pointed out that triples \mbox{2He + H} constitute about 65\% among 50 "white" stars found to that time. However, origin of this effect has not been studied being in a "shadow" of emerging studies with radioactive nuclei. Meanwhile, the the \mbox{2He + H} triple dominance indicate the possible presence in ${}^{10}$B of structures \mbox{${}^{9}$B$_{g.s.}$ + $\mit{n}$} side by side with the mirror one \mbox{${}^{9}$Be + $\mit{p}$}. It is interesting to verify whether they have equal contributions or not. Another opportunity is that the ${}^{10}$B nucleus can incorporate the "dilute" ${}^{9}$Be cluster in the superpositions \mbox{${}^{8}$Be$_{g.s.}$ + $\mit{n}$} and \mbox{${}^{8}$Be$_{2^+}$ + $\mit{n}$}. Both them are leading to 3-prong "white" stars out of ${}^{9}$B$_{g.s.}$ decays. Thus, a new round of the ${}^{10}$B analysis is started recently which progress is summarized below.

A significant increasing statistics of stars \mbox{${}^{10}$B $\to$ 2He + H} is reached in an accelerated search for paired tracks when scanning is performed along transverse strips of NTE layers. Early, such an approach allowed one to obtain statistics of 500 events \mbox{${}^{9}$B $\to$ 2He} in a reasonable labour time. To date, measurements of emission angles of relativistic fragments are performed in 297 events \mbox{${}^{10}$B $\to$ 2He + H} including 143 "white" stars (Fig. \ref{fig:Fig.12}). 

The distribution of 2He pairs over the opening angle $\Theta_{2He}$ in an interval \mbox{0 $<$ $\Theta_{n(arrow)}$ $<$ 10.5 mrad} allows one to count 57 decays ${}^{8}$Be$_{g.s.}$ in all found events \mbox{${}^{10}$B $\to$ 2He + H} including 36 in the "white" stars (Fig. \ref{fig:Fig.7}, right). These numbers give \mbox{19 $\pm$ 3\%} and \mbox{25 $\pm$ 5\%} in the respective statistics. Then, the condition on the opening angle \mbox{$\Theta_{{}^{8}Be_{g.s.} + H}$ $<$ 25 mrad} (Fig. \ref{fig:Fig.9}, right) allows one to identify 30 decays ${}^{9}$B in all found events and 20 in the "white" stars which constitute, respectively, \mbox{10 $\pm$ 2\%} and \mbox{14 $\pm$ 3\%} contributions of the subset \mbox{${}^{8}$Be$_{g.s.}$ + H} in the channel \mbox{2He + H}. Thus, in the "white" star case decays ${}^{9}$B explains just \mbox{56 $\pm$ 16\%} of decays ${}^{8}$Be$_{g.s.}$. This way, the idea about simultaneous coexistence in ${}^{10}$B of superposition of cores ${}^{8}$Be$_{g.s.}$, ${}^{8}$Be$_{2^+}$  and ${}^{9}$B obtain a ground.

Statistics of "white" stars found without sampling (Table. \ref{tabular:Tab.2}) one allows to compare the probability of dissociation in the channels \mbox{${}^{9}$B + $\mit{n}$} and \mbox{${}^{9}$Be + $\mit{p}$} (two events). Measurements of fragment emission angles have become possible only in 65 of the 108 events \mbox{2He + H}, which determines the reconstruction efficiency of the eight ${}^{9}$B decays found among them. On this basis, a probability ratio of the mirror channels \mbox{${}^{9}$B + $\mit{n}$} and \mbox{${}^{9}$Be + $\mit{p}$} is estimated to be \mbox{6 $\pm$ 1}. Accounting for observation efficiency of the "white" stars \mbox{${}^{9}$Be + $\mit{p}$} does not affect qualitatively this ratio.

This fact is quite unexpected and even intriguing. Perhaps it points to the predominance of the ${}^{9}$Be core in nuclear molecular form \mbox{2$\alpha$ + $\mit{n}$} appearing in the dissociation channels containing ${}^{8}$Be$_{2^+}$ or ${}^{8}$Be$_{g.s.}$ without ${}^{9}$B decays. The core ${}^{9}$B represents such a structure originally. Another explanation may be based on a broader spatial distribution of neutrons in the ${}^{10}$B compared to protons.

The distributions of the energy of \mbox{$\alpha$-particle} pairs $\mit{Q}_{2\alpha}$ and \mbox{2$\alpha + p$} triples $\mit{Q}_{2\alpha p}$ from the found \mbox{${}^{10}$B $\to$ 2He + H} events shown in Fig. \ref{fig:Fig.12} arrange a common ensemble with the considered cases ${}^{9}$Be, ${}^{10}$C and ${}^{11}$C. Correct positioning of the peaks ${}^{8}$Be$_{g.s.}$ and ${}^{9}$B can be noted. Thus, evolution of structural changes related with the unstable nuclei obtains an experimental extended ground. 

\begin{figure}[t]
	\centerline{\includegraphics*[width=0.75\linewidth]{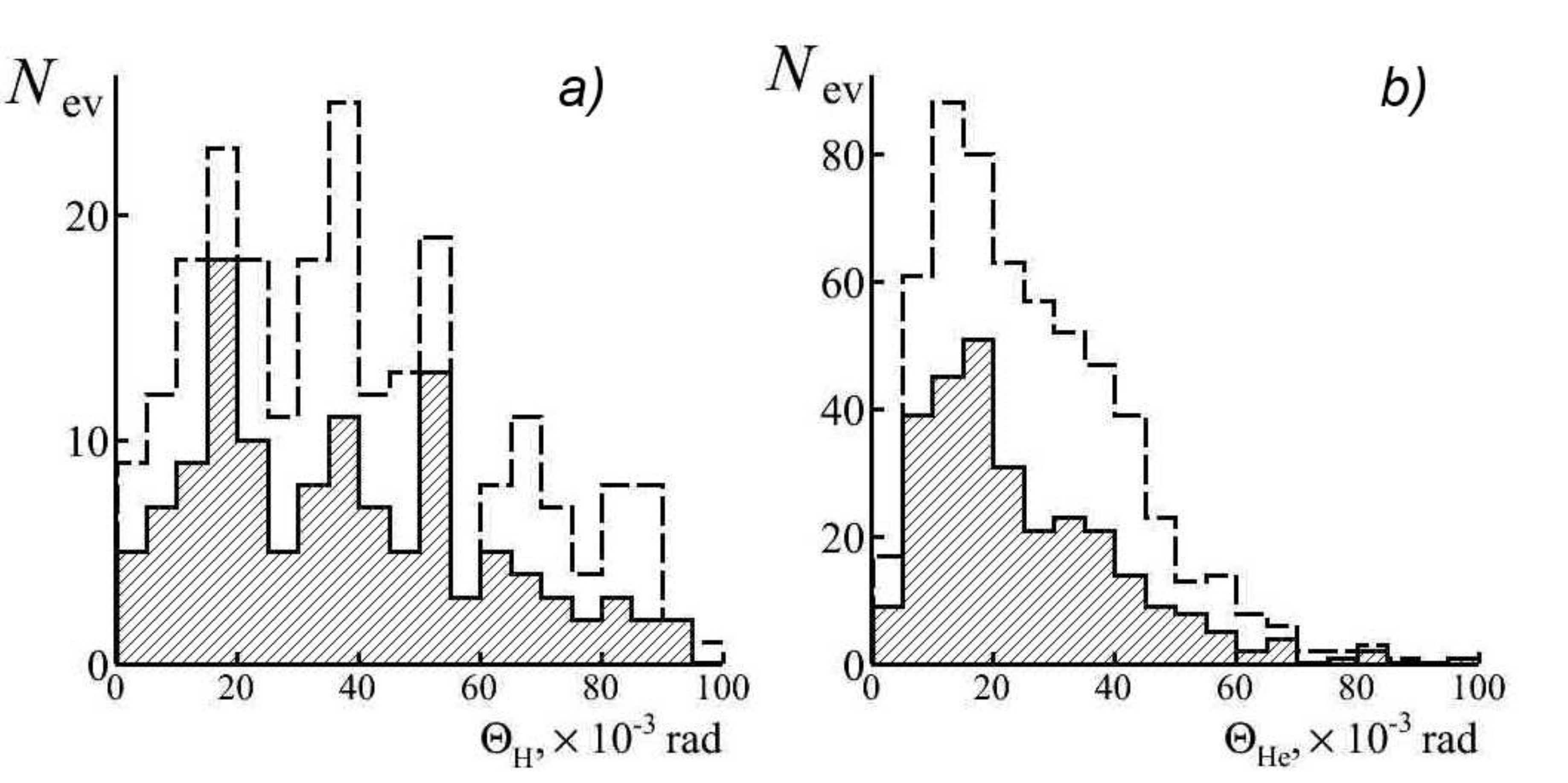}}
	\caption{Distributions for channel 2He + H over emission angles of fragments $\Theta_{He}$ and $\Theta_{H}$ in all found ${}^{10}$B stars (a, dashed) and ${}^{10}$B "white"  stars (a, hatched).}
	\label{fig:Fig.11}
\end{figure}

\begin{figure}[t]
	\centerline{\includegraphics*[width=0.75\linewidth]{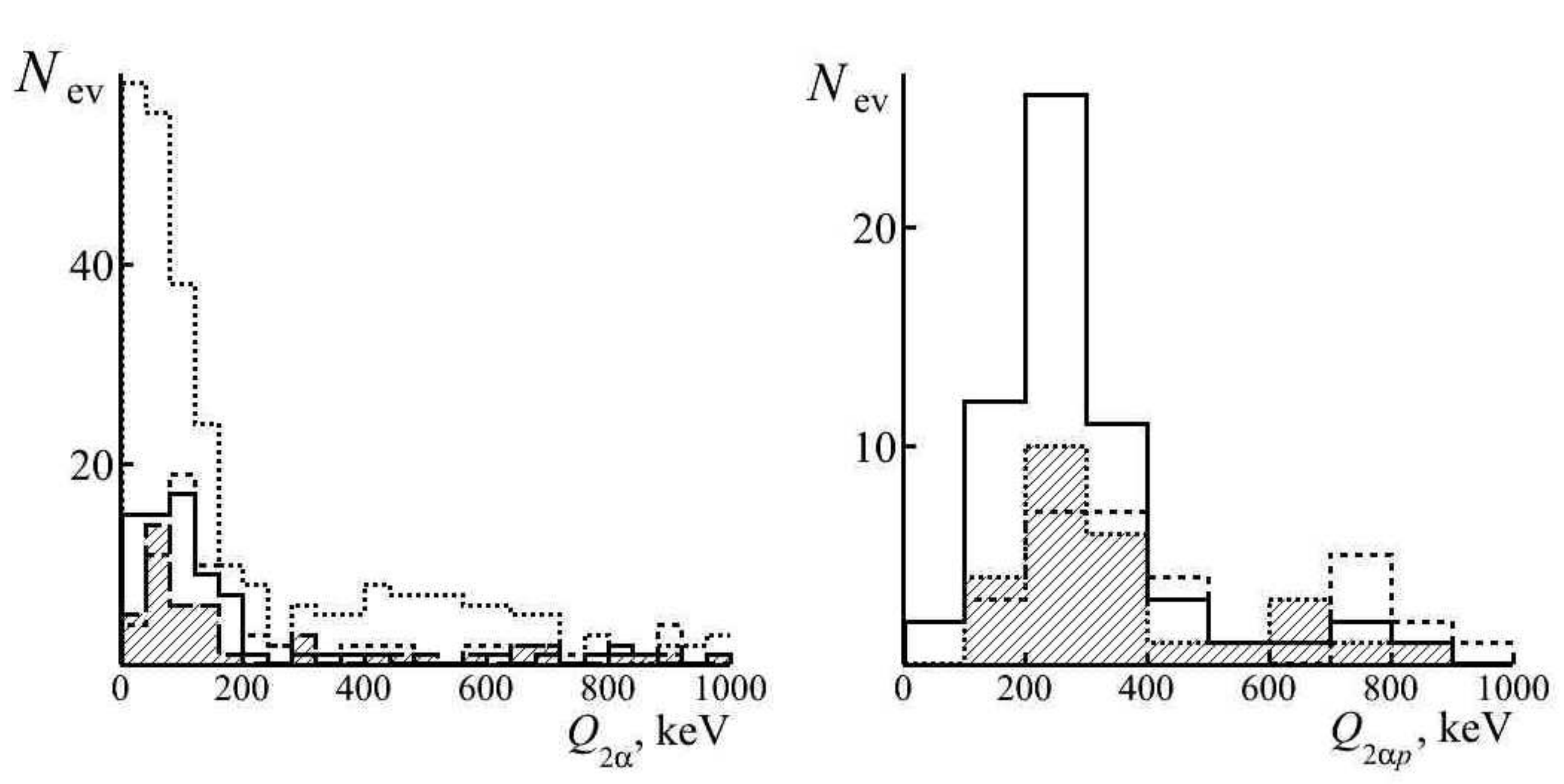}}
	\caption{Distributions of all found stars ${}^{9}$Be $\to$ 2He (left top, dotted), \mbox{${}^{10}$B $\to$ 2He + H}, \mbox{${}^{11}$C $\to$ 2He + 2H} (hatched) and "white" stars \mbox{${}^{10}$C $\to$ 2He + 2H} (solid) over energy $\mit{Q}_{2\alpha}$ of 2$\alpha$ pairs and energy $\mit{Q}_{2\alpha p}$ of 2$\alpha p$ triples.}
	\label{fig:Fig.12}
\end{figure}

Identification of He and H isotopes by a multiple scattering method progressing now will promote the analysis. In particular, the cluster configuration involving the deuteron \mbox{${}^{8}$Be$_{2^+}$ + $\mit{d}$} can be a source of ${}^{8}$Be$_{2^+}$ decays. Besides, since the channel \mbox{${}^{10}$B $\to$ ${}^{6}$Li + $\alpha$} is observed with 10\% probability contribution of the "dilute" ${}^{6}$Li cluster into the \mbox{2$\alpha + p(d)$} channel can be expected. Thus, with attraction of existing knowledge on ${}^{9}$Be and ${}^{6}$Li the pattern ${}^{10}$B dissociation via decays ${}^{8}$Be$_{g.s.}$, ${}^{8}$Be$_{2^+}$ and ${}^{9}$B can be disentangled step by step. If successful, it will lead to better understanding for the neighboring nuclei ${}^{11}$C and, then, ${}^{12}$N.

\section*{Summary}

\noindent Contribution of the unstable nuclei ${}^{6}$Be, ${}^{8}$Be and ${}^{9}$B into dissociation of relativistic nuclei ${}^{7,9}$Be, ${}^{10}$B and ${}^{10,11}$C is under study on the basis of the nuclear track emulsion exposed to secondary beams of the JINR Nuclotron. 

On the basis of angular measurements \mbox{27 $\pm$ 5\%} of events \mbox{${}^{7}$Be $\to$ He + 2H} can be attributed to ${}^{6}$Be decays. Contribution of the configuration \mbox{${}^{6}$Be + $\mit{n}$} to the ${}^{7}$Be structure is estimated at a level of \mbox{8 $\pm$ 1 \%} which is near the value of \mbox{5 $\pm$ 1\%} for the configuration \mbox{${}^{6}$Li + $\mit{p}$}.

Distributions over the opening angle of \mbox{$\alpha$-pairs} indicate to a simultaneous presence of virtual ${}^{8}$Be$_{g.s.}$ and ${}^{8}$Be$_{2^+}$ states in the ground states of the ${}^{9}$Be and ${}^{10}$C nuclei. 

The core ${}^{9}$B is manifested in the ${}^{10}$C nucleus with a probability of \mbox{30 $\pm$ 4\%}.  ${}^{8}$Be$_{g.s.}$ decays in ${}^{10}$C "white" stars always arise through the ${}^{9}$B decays. For ${}^{10}$C "white" stars it have to be assumed that ${}^{6}$Be and ${}^{8}$Be$_{g.s.}$ are produced as interfering parts of $2\alpha2p$ ensembles due to impossibility of separation of the ${}^{6}$Be and ${}^{8}$Be$_{g.s.}$ decays. Selection of the ${}^{10}$C "white" stars accompanied by ${}^{8}$Be$_{g.s.}$ (${}^{9}$B) leads to appearance in the excitation energy distribution of $2\alpha2p$ "quartets" of the distinct peak with a maximum at \mbox{4.1 $\pm$ 0.3} MeV.

In a charge state distribution of fragments the share of the channel ${}^{10}$B $\to$ 2He + H is 77\%. On the basis of measurements of fragment emission angles it is determined that unstable nucleus ${}^{8}$Be$_{g.s.}$ manifests itself with a probability of \mbox{25 $\pm$ 5\%} where \mbox{14 $\pm$ 3\%} of them occur in decays of the unstable nucleus ${}^{9}$B. Channel \mbox{Be + H} appeared subdued accounting for about 2\% of "white" stars. A probability ratio of the mirror channels \mbox{${}^{9}$B + $\mit{n}$} and \mbox{${}^{9}$Be + $\mit{p}$} is estimated to be \mbox{6 $\pm$ 1}.

${}^{8}$Be$_{g.s.}$ decays are presented in \mbox{24 $\pm$ 7\%} of \mbox{2He + 2H} and \mbox{27 $\pm$ 11\%} of the 3He of the ${}^{11}$C "white" stars. ${}^{9}$B decays are identified in "white" stars \mbox{${}^{11}$C $\to$ 2He + 2H} constituting 14\% of the ${}^{11}$C "white" stars. As in the ${}^{10}$C case ${}^{8}$Be$_{g.s.}$ decays in ${}^{11}$C "white" stars almost always arise through ${}^{9}$B decays. On this ground the channel \mbox{${}^{9}$B + H} amounts \mbox{14 $\pm$ 3\%}.	

It should be noted that for the nuclei ${}^{11}$C and ${}^{12}$N comes into play restriction of our approach based on coherent dissociation of relativistic nuclei in NTE consisting in the inability of a direct identification of mass numbers of relativistic fragments heavier than He. Shares of events with participation of such fragments should increase rapidly with increasing mass number of a nucleus under study. This identification is possible in electronic experiments with magnetic analysis in a range of energy of a few GeV per nucleon of beam nuclei. Studies using the NTE technique will keep the value for orientation of experiments on coherent dissociation of relativistic \mbox{neutron-deficient} nuclei. In perspective, identification is possible at energy of nuclei in the region of tens of GeV per nucleon in experiments with hadron calorimeters.

In conclusion the authors are grateful to their colleagues A. I. Malakhov, K. Z. Mamatkulov, R. R. Kattabekov in Veksler\&Baldin Laboratory of High Energy Physics of JINR and Sergei Petrovich Kharlamov, their senior comrade in the Lebedev Physical Institute for cooperation and critical discussions related with this review.


\newpage
\begin{thebibliography}{}
	
	\bibitem{1}``The BECQUEREL Project''. WEB site: \href{http://becquerel.jinr.ru/text/books/POWELL.pdf}{http://becquerel.jinr.ru/.}; 
	
	\bibitem{2} C. F. Powell, P. H. Fowler, and D. H. Perkins  "Study of Elementary Particles by the Photographic Method",  Pergamon, London, 1959;
	\bibitem{3} W. H. Barkas "Nuclear Research Emulsions" Academic Press, New York , London, 1963.
	\bibitem{4} Y. Goldschmidt-Cremont "Photographic Emulsions" Annu. Rev. Nucl. Sci. 1953;
	\bibitem{5}  \href{http://link.springer.com/chapter/10.1007\%2F978-3-319-01077-9_3}{P. I. Zarubin "Tomography" of the cluster structure of light nuclei via relativistic dissociation" Lecture Notes in Physics 875, Clusters in Nuclei, vol. 3, 51-93 (2014) Springer International Publishing};  \href{http://search.arxiv.org:8081/paper.jsp?r=1309.4881&qid=1469043284309bas_nCnN_1074364797&qs=1309.4881}{arXiv: 1309.4881}; 
	\bibitem{6}  \href{http://link.springer.com/article/10.1134\%2FS0021364008140014}{N. G. Peresadko \textit{et al} "Role of the Nuclear and Electromagnetic Interactions in the Coherent Dissociation of the Relativistic ${}^{7}$Li Nucleus into the \mbox{${}^{3}$H + ${}^{4}$He} Channel" Phys. At. Nucl. 88, 75-79 (2008)};
	\bibitem{7}   \href{http://link.springer.com/article/10.1134/S1063778810110177}{N. G. Peresadko \textit{et al} "Fragmentation of ${}^{7}$Li Relativistic Nuclei on a Proton into the \mbox{${}^{3}$H + ${}^{4}$He} Channel" Phys. At. Nucl. 73, 1942-1947 (2010)};
	\bibitem{8}   \href{http://link.springer.com/article/10.1007/s00601-014-0832-4}{N. K. Kornegrutsa \textit{et al} "Clustering features of the ${}^{7}$Be nucleus in relativistic fragmentation" Few Body Syst. 55 1021-1023 (2014)} ;  \href{http://arxiv.org/abs/1410.5162}{arXiv: 1410.5162};
	\bibitem{9}     \href{http://link.springer.com/article/10.1134\%2FS1063778815020052}{Yu. A. Alexandrov \textit{et al} "Dissociation of Relativistic ${}^{7}$Be Nuclei through the ${}^{3}$He + ${}^{4}$He Channel on a Proton Target" Phys. At. Nucl. 78, 363-368 (2015)};
	\bibitem{10}  \href{http://link.springer.com/article/10.1134/S1063778813100141}{K. Z. Mamatkulov \textit{et al} "Dissociation of ${}^{10}$C Nuclei in a Track Nuclear Emulsion at Energy of 1.2 GeV per Nucleon" Phys. At. Nucl. 76, pp. 1224-1229 (2013)} ;  \href{http://arxiv.org/abs/1309.4241}{arXiv: 1309.4241};
	\bibitem{11} \href{http://link.springer.com/article/10.1134\%2FS1063778813100074}{ R. R. Kattabekov \textit{et al} "Coherent Dissociation of Relativistic ${}^{12}$N Nuclei" Phys. At. Nucl. 76, 1219-1223 (2013)};  \href{http://arxiv.org/abs/1310.2080}{arXiv: 1310.2080};
	\bibitem{12}  \href{http://link.springer.com/article/10.1134\%2FS1063778815060022}{D. A. Artemenkov \textit{et al} "Charge topology of coherent dissociation of ${}^{11}$C and ${}^{12}$N relativistic nuclei" Phys. At. Nucl. 78, 794-799 (2015)} ;  \href{http://arxiv.org/abs/1411.5806}{arXiv: 1411.5806};
	\bibitem{13} \href{http://slavich.ru}{"Slavich Company JSC". WEB site: www.slavich.ru}, \href{http://newslavich.com}{www.newslavich.com};
	\bibitem{14}   \href{http://link.springer.com/article/10.1134/S1547477113050026}{D. A. Artemenkov \textit{et al} "Exposure of Nuclear Track Emulsion to ${}^{8}$He Nuclei at the ACCULINNA Separator" Phys. Part. Nucl. Lett. 10, 415-421(2013)}; \href{http://arxiv.org/abs/1309.4808}{arXiv: 1309.4808};
	\bibitem{15}  \href{http://link.springer.com/article/10.1007/s00601-014-0885-4}{D. A. Artemenkov \textit{et al} "${}^{8}$He nuclei stopped in nuclear track emulsion" Few-Body Syst. 55, 8-10, 733736 (2014)};  \href{http://arxiv.org/abs/1410.5188}{arXiv: 1410.5188};
	\bibitem{16}   \href{http://www.epj-conferences.org/articles/epjconf/pdf/2014/03/epjconf_inpc2013_11044.pdf}{P. I. Zarubin\textit{et al} "${}^{8}$He nuclei stopped in nuclear track emulsion" EPJ, \mbox{Web of Conf. 66, 11044, 2014};}
	\bibitem{17}   \href{http://arxiv.org/abs/1407.4575}{R. R. Kattabekov \textit{et al}  "Correlations of $\alpha$-particles in splitting of ${}^{12}$C nuclei by neutrons of energy of 14.1 MeV" Phys. At. Nucl. 76, add. issue (Russian) 88-91(2013); arXiv: 1407.4575};
	\bibitem{18} \href{http://link.springer.com/article/10.1134\%2FS106377881504002X}{ D. A. Artemenkov \textit{et al} "Irradiation of Nuclear Track Emulsions with Thermal Neutrons, Heavy Ions, and Muons"  Phys. At. Nucl. 78, 579-585 (2015)} ; \href{http://arxiv.org/abs/1407.572}{arXiv: 1407.572};
	\bibitem{19} \href{http://link.springer.com/chapter/10.1007/978-3-319-01077-9_6}{D. V. Kamanin, Y. V. Pyatkov "Clusterization in ternary fission" 875, Clusters in Nuclei, vol. 3 184-246 (2013), Springer International Publishing (and references herein)};
	\bibitem{20} \href{http://www.sciencedirect.com/science/article/pii/S1875389215014066}{K. Z. Mamatkulov \textit{et al} "Toward an automated analysis of slow ions in nuclear track emulsion" Phys. Procedia 74, 59-66 (2015)}; \href{http://arxiv.org/abs/1508.2707}{arXiv: 1508.2707}.
	
	
	
	
\end{thebibliography}
\end {document}